
\input amstex
\documentstyle{amsppt}
\hoffset=.1in

\define\a{\alpha}
\define\be{\beta}

\define\de{\delta}

\define\vd{\varDelta}
\define\vl{\varLambda}

\define\cb{\Bbb C}

\define\rr{\Bbb R}

\define\cad{\Cal D}
\define\cah{\Cal H}

\define\cam{\Cal M}

\define\pti{\tilde p}
\define\rti{\tilde R}
\define\vti{\tilde v}
\define\lti{\tilde \varLambda}

\define\ham{\hat M}

\define\pd#1#2{\dfrac{\partial#1}{\partial#2}}

\define\vc#1{(#1_1,\ldots,#1_n)}
\define\vct#1{[#1_1,\ldots,#1_n]}
\define\vect#1{\{#1_1,\ldots,#1_n\}}


\define\bs{\backslash}

\define\ptak{e^{imv^0}}
\define\sowa{e^{-imv^0}}
\define\deja{\Cal D_{ij}}
\define\deka{\Cal D_{kj}}
\define\jak{\frac{i}{\kappa}}
\define\mak{\left(\frac{m}{\kappa}\right)}

\define\1{Poincare group}
\define\2{unitary}
\define\3{quantum}
\define\4{group}
\define\5{commut}
\define\6{representation}
\define\7{character}
\define\8{coequivariance}
\define\9{general}
\define\0{automorphism}

\document

\topmatter
\title The induced representations of the $\kappa$-Poincare
group.\\
 The massive case
\endtitle
\rightheadtext{$\kappa$-Poincare group}
\author Pawe\l\  Ma\'slanka*\\
{\it Department of Functional Analysis}\\
{\it Institute of Mathematics, University of \L\'od\'z}\\
{\it ul. St. Banacha 22, 90--238 \L\'od\'z, Poland}
\endauthor
\leftheadtext{Pawe\l\ Ma\'slanka}

\thanks
*\ Supported by KBN grant 2 0218 91 01
\endthanks

\abstract The induced \6s of the $\kappa$-\1 for the massive case
are described. It is shown that it extends many of the features
of the classical case.
\endabstract
\endtopmatter
\document
\head 1. Introduction
\endhead

Recently an interesting deformation of Poincare algebra - the
so-called $\kappa$-Poincare algebra - has been constructed [1]
(see also [2]). Its \7istic property is that the deformation
parameter $\kappa$ is dimensionful. Some physical consequences of
the deformed space-time symmetry have been discussed by H. Bacry
[3] (see also [4]) who stressed its attractive features. The
global counterpart of the $\kappa$-Poincare algebra was
constructed by S. Zakrzewski [5]. The resulting \3 \1 can be
described as follows. The group element is written in the
following form
$$
g = \bmatrix
\vl & v\\
0 & 1
\endbmatrix
\tag{1}
$$
where $\vl = [\vl^\mu \nu]_{\mu,\nu = 0}^3$ and $v = [v^\mu]_{\mu =
0}^3$ are selfadjoint elements subject to the following \5ation
rules
$$
\aligned
[ \vl^\mu_\nu, \vl^\a_\be]& = 0,\\
[v^r,v^k]& = 0,\\
[v^0,v^r]&= -\jak v^r,\\
[\vl^\mu \nu,v^0]& = \jak (\vl^\mu_0 \vl^0_\nu - \de^{\mu
0}\de_{\nu 0}),\\
[ \vl^m_0,v^r] &= \jak(\vl^m_0 \vl^r_0 + \de_{mr} - \de_{mr}\vl^0_0),\\
[\vl^0_m,v^r]&= \jak(\vl^0_0 \vl^r_m - \vl^r_m),\\
[\vl^m_n,v^r]&= \jak(\vl^m_0 \vl^r_n - \de_{mr} \vl^0_n),\\
[\vl^0_0,v^r]&= \jak(\vl^r_0 \vl^0_0 - \vl^r_0).
\endaligned
\tag{2}
$$
Equipped with the standard matrix comultiplication the above
structure  defines a Hopf $*$-algebra.

Once the \3 \1 is defined we can try to explore its physical
consequences. The necessary step in this direction is to
construct its unitary \6s, to define its action on \3 Minkowski
space, to find covariant fields, etc. In this paper we
concentrate on the first problem: to find the \2 \6s. To this end
we shall use the method of induced \6s for \3 \4s as formulated
by A. Gonzales-Ruiz and L.A. Ibort [6].

\head II. Some subgroups
\endhead

To start with let us analyse some sub\4s of \3 $\kappa$-\1
defined above. Let us recall that, by definition, $A(K)$ is a \3
sub\4 of a \3 \4 $A(G)$ if there exists an epimorphism of Hopf
algebras $\Pi : A(G) \to A(K)$; we have
$$
(\Pi \otimes \Pi) \circ \vd_G = \vd_K \circ \Pi.
\tag{3}
$$
Let $A(G)$ be our \3 \1. We are interested in those sub\4s $A(K)$
for which $\Pi(v^\mu)$ are independent generators. First let us
note that, due to the first \5ation rule (2) and the form of the
coproduct, $\{\Pi(\vl^\mu_\nu)\}_{\mu,\nu = 0}^3$ generate a sub\4
(in the classical sense) of the classical Lorentz \4. There are
few choices of this sub\4 which lead to the \3 sub\4s of \3
\1.
\roster
\item"{(i)}" Let us first take $\{\Pi(\vl^\mu_\nu)\}$ to be a
trivial sub\4 of the Lorentz \4. Then $A(N)$ is generated by four
elements $\vti^\mu$ such that
$$
\aligned
[\vti^0,\vti^k] & = - \jak \vti^k,\\
[\vti^i,\vti^k] & = 0,\\
\vd(\vti^\mu) =  \vti^\mu  \otimes  I + I \otimes \vti^\mu.
\endaligned
\tag{4}
$$
The epimorphism $\Pi$ is given by
$$
\Pi ( v^\mu) = \vti^\mu, \qquad \Pi(\vl^\mu_\nu) =
\de^\mu_\nu I.
\tag{5}
$$
\item"{(ii)}" Let $\{\Pi(\vl^\mu_\nu)\}$ be the rotation sub\4.
Define $A(K)$ as generated by $\vti^\mu$ and $M^i_j$ obeying:
$$
\aligned
&[\vti^0,\vti^k]  = - \jak \vti^k, \qquad [\vti^i,\vti^k]  = 0,\\
&[M^i_j,M^k_l] = 0, \qquad [M^i_j,\vti^\mu] = 0,\\
& M^i_j M^k_j = \de_{ik}I, \qquad M^j_l M^j_k = \de_{ik}I,\\
& \vd(M^i_j) = M^i_k \otimes M^k_j,\\
&\vd (\vti^i) = M^i_k \otimes \vti^k + \vti^i \otimes I,\\
& \vd(\vti^0) = I \otimes \vti^0 + \vti^0 \otimes I.
\endaligned
\tag{6}
$$
Epimorphism $\Pi$ is defined as follows
$$
\aligned
& \Pi(v^\mu) = \vti^\mu, \qquad \Pi(\vl^0_0) = I,\\
&\Pi(\vl^k_0) = \Pi(\vl_k^0) = 0, \qquad \Pi(\vl^i_j) = M^i_j.
\endaligned
\tag{7}
$$
It is easy to check that $A(K)$ is a sub\4 of \3 \1.
\endroster

We shall not discuss here which sub\4s of (classical) Lorentz \4
generate (in the sense explained above) \3 sub\4s of $\kappa$-\1.
Let us only note two interesting features. First, the whole
Lorentz \4 does not form a sub\4 of \3 \1. Moreover, the same
continues to hold for those sub\4s of Lorentz \4 which are the
stability \4s of light-like or space-like fourvectors. To see
this let us first note that space rotations are the \0s of our
\1. Indeed, if $R$ is any $c$-number $3\times 3$ orthogonal
matrix and
$$
\rti = \bmatrix
1 & 0\\
0 & R
\endbmatrix
$$
then
$$v^0 \to v^0, \qquad v^i \to (Rv)^i, \qquad \vl \to \rti \vl
\rti^{-1}
$$
is an \0 of $\kappa$-\1. Therefore we can put the light-like
(resp. space-like) fourvector in canonical position $k^\mu =
(k,0,0,k)$ (resp. $k^\mu = (0,0,0,m)$. Now assume that
$\lti^\mu_\nu \equiv \Pi(\vl^\mu_\nu)$ is a stability \4 of, say,
light-like fourvector $k^\mu$. Then $\lti^\mu_\nu k^\nu = k^\mu$
implies
$$
\lti^0_0 + \lti^0_3 = I, \qquad \lti^3_0 + \lti^3_3 = I, \qquad
\lti^{1,2}_0 + \lti^{1,2}_3 = 0;
\tag{8}
$$
these are the standard conditions. However, further constraints
follow from the \5ation rules (2) and the fact that $\Pi$ is a
homomorphism. Taking the \5ators of both sides of eqs. (8) with
$\vti^\mu \equiv \Pi(v^\mu)$ we arrive finally at the following
form of $\lti^\mu_\nu$:
$$
\lti =
\bmatrix
1 & 0 & 0 & 0\\
0 & \cos \a & -\sin \a & 0\\
0 & \sin \a & \cos \a & 0\\
0 & 0 & 0 & 1
\endbmatrix
$$
We see that, contrary to the classical case, the stability sub\4
of light-like fourvector does not give rise to the sub\4 of
$\kappa$-\1 (in the sense defined above). However, the
$U(1)$-sub\4 of the stability \4 does the job. The same holds
true for the space-like case: again only rotations around the
third axis survive in the \3 case.

\head III. Representations
\endhead
Let us now construct the induced \6s for the massive case, i.e.
we choose the \3 sub\4 described in (ii). In order to describe
explicitly these \6s we could follow [6]. In this paper the \3
counterpart of the classical construction of induced \6s is
given. As it is well known [7] in order to construct the \6s of a
\4 $G$ induced from the \6s of its sub\4 $H \subset G$ one starts
from the Hilbert space of square integrable functions on $G$
taking their values in the vector space carrying some \6 of $H$.
The \4 action is defined to be, say, right action: $f(g) \to
f(gg_0)$. The essence of the method is the selection of invariant
subspace by imposing the so-called \8 condition; in many cases the
invariant subspace obtained in this way carries an irreducible \6
of $G$. The whole construction can be \9ized in a rather
straightforward way to the \3 case [6]. However, in specific
cases it is profitable to have more explicit \7ization of the \6.
This is achieved by solving explicitly the \8 condition which
gives rise to the description of \6 in terms of Hilbert space of
functions defined on the right coset space $H \bs G$. This is
especially effective for the case of semidirect products in which
one factor $N$ is abelian. Then the irreducible \6s are induced
from the stability sub\4s of \7s of $N$ and the relevant coset
spaces get with relevant orbits in the dual \4 $\hat N$. In
particular, in the case of \1 we obtain exactly Wigner's
construction.

Let us now apply the same idea to the case of $\kappa$-\1. We
start with abelian sub\4 of translations (our counterpart of $N$)
described in (ii). It is abelian in the sense that it is \5ative.
We define its dual in a similar way as in the classical case.
Recall that a \6 $\rho$ of a \3 \4 $A(N)$ is a map $\rho : \cah
\to A(N) \otimes \cah$ satisfying $(\vd_N \otimes I)\circ \rho =
(I \otimes \rho ) \circ \rho$ (or, for the right (co)\6: $\rho : \cah
\to \cah \otimes A(N)$, $(I \otimes \vd_N)\circ \rho =
(\rho \otimes I ) \circ \rho$). If $\cah$ is onedimensional, an
\2 \6 $\rho$ can be written as:
$$
\rho : z \to a \otimes z, \qquad z \in \cb,
$$
where $a$ is \2 element of $A(N)$ and
$$
\vd_N(a) = a \otimes a.
\tag{9}
$$
The \2 elements of our algebra $A(N)$ obeying (9) are called \7s.
It is trivial to verify that a product of \7s is again a \7. It
is not difficult to check that in our case the solution to eq.
(9) reads
$$
a = e^{iq_0\vti_0} e^{iq_k\vti_k}
\tag{10}
$$
where $q_0$, $q_k$ are real numbers. Now, if $a'$ is another
character, a small calculation gives
$$
aa' = e^{iq_0''\vti_0} e^{iq_k''\vti_k}
\tag{11a}
$$
where
$$
q_0'' = q_0 + q_0', \qquad q_k'' = q_ke^{-\frac{q_0'}{\kappa}} + q_k'.
\tag{11b}
$$
To get slightly more symmetric form we redefine
$$
p_0 \equiv q_0, \qquad p_k = q_k e^{q_0\slash 2\kappa}.
\tag{12}
$$
Then we conclude that the dual \4 $\hat N$ is the classical \4
with \4 manifold being $\rr^4$ and the composition law
$$
\{p_0,p_k\}\ \ast \ \{p_0',p_k'\} = \left\{p_0 + p_0',\,
p_ke^{-\frac{p_0'}{2\kappa}} + p_k'
e^{\frac{p_0}{2\kappa}}\right\}.
\tag{13}
$$
The only difference with the classical case is that the dual \4
is no longer abelian but only solvable. However, we shall see
that our space of states can still be viewed (in some sense) as
consisting of functions concentrated on some orbit in $\hat N$
and taking values in the space carrying a \6 of $A(K)$.

Let us now construct the \6s of $A(K)$. We shall follow closely
the classical case. There the \6 of $A(K)$ from which an
irreducible \6 of \1 is induced is constructed out of \6 of
rotation sub\4 and a \7 of translation sub\4 invariant under the
action of the former (cf. [7]). It is easy to see that, in \3
case, both rotations\footnote"*)"{ \ note that rotations form a
classical group} and translations also form the sub\4s of $A(K)$;
by direct analysis of the classical case we infer also that the
counterpart of an invariant \7 is here provided by a \2 element
$a$ constructed out of $v^\mu$ and obeying
$$
\vd_K(a) = a\otimes a.
\tag{14}
$$
Taking the above into account we construct the \6 of $A(K)$ as
follows. Let $\cah$ be a Hilbert space carrying unitary \6 of
classical rotation \4
$$
\cah \ni \psi = a_ie_i \to \deja(M) a_ie_j \in \cah.
\tag{15}
$$
Then $\rho : \cah \to \cah \otimes A(K)$ is defined as follows:
$$
\rho (\psi) \equiv \rho( a_ie_i) =  a_ie_j \otimes \deja(M) \ptak
\tag{16}
$$
where $m$ is a numerical parameter, $m \in \rr^+$, and $\{e_i\}$
is the basis in $\cah$.

It is easy to check that $\rho$ is really a \6. We have
$$
[(\rho \otimes I) \circ \rho](\psi) = a_ie_k \otimes \deka(M)
\ptak \otimes \deja(M) \ptak
\tag{17a}
$$
and, on the other hand,
$$
\aligned
[(I \otimes \vd_K) \circ \rho](\psi) & = a_ie_j \otimes \deja(M
\otimes M) e^{im(v^0 \otimes I + I \otimes v^0)}\\
& = a_i e_j \otimes (\deka (M) \otimes D_{ki}(M))( \ptak \otimes
\ptak). \endaligned
\tag{17b}
$$
Note that the property (14) of $a \equiv \ptak$ has played a
crucial role above.

Now, we are ready to solve the \8 condition explicitly. Let us
recall that the Hilbert space of the induced \6 of \3 \4 $(A(G))$
is a subspace of $\cah \otimes A(G)$ consisting of the elements
subject to the following \8 condition:
$$
\cah \uparrow A(G) = \{F \in \cah \otimes A(G) : (I \otimes L) F
= (\rho \otimes I)F\}.
\tag{18}
$$
Here $L$ is a left coaction of $A(K)$ in $A(G)$ [6]:
$$
L = (\Pi \otimes I) \circ \vd_G.
\tag{19}
$$
To solve the equation $(I \otimes L) F = (\rho \otimes I)F$ let
us first make what can be called a counterpart of Mackey
decomposition
$$
\bmatrix
\vl & v\\
0 & 1
\endbmatrix
=
\bmatrix
\ham & v\\
0 & 1
\endbmatrix
\bmatrix
\lti & 0\\
0 & 1
\endbmatrix
\tag{20}
$$
where the product on the right-hand side is an algebra product
(and not the tensor one). The matrices $\ham$ and $\lti$ are
defined as follows
$$
\aligned
\lti & =
\bmatrix
\vl^0_0  & \vl_j^0\\
\vl^0_i  & \de_{ij}I + \frac{\vl^0_i \vl^0_j}{1 + \vl^0_0}
\endbmatrix \\
\ham & = \bmatrix
1 & 0 \\
0 & \cam
\endbmatrix
\endaligned
\tag{21}
$$
where
$$
\cam = [\cam^i_j]_{i,j = 1}^3 = \left[ \vl_j^i - \frac{\vl^i_0
\vl^0_j}{1 + \vl^0_0}\right]_{i,j = 1}^3.
$$
Note that the matrix $\ham$, being formally orthogonal, is not an
element of the subalgebra $A(K)$, in particular, their elements
do not commute with $v^{k'}s$. If we define the momenta $p_\mu$
by
$$
p_\mu = m \vl^0_\mu
\tag{22}
$$
we can write $\lti$ as
$$
\lti = \bmatrix
\frac{p_0}{m} & \frac{p_j}{m} \\
\frac{p_i}{m} & \de_{ij}I + \frac{p_ip_j}{m(p_0 + m)}
\endbmatrix
\tag{23}
$$
Now, we propose the following solution to the \8 condition
$$
F = e_i \otimes \deja(\cam)\ptak  f_j (\lti  )\equiv e_i \otimes
\deja(\cam) \ptak f_j(p).
\tag{24}
$$
The matrices $\deja(\cam)$ are constructed out of $\cam$'s in the
same way as the matrices of the \6 of classical orthogonal \4.
There is no ambiguity here because all elements of $\cam$ \5e
among themselves. Note that, in principle, we should take only
integer spin \6s because we have quantized the \1 and not
$ISL(2,\cb)$.

Now, we can check that our Ansatz solves the \8 condition. One
has
$$
(\rho \otimes I)(F) = e_k \otimes \cad_{ki}(M) \ptak \otimes
\deja(\cam) \ptak f_j(\lti).
\tag{25}
$$
On the other hand,
$$
(I \otimes L)(F) = e_k \otimes \{ (\Pi \otimes I)(\deka
(\vd(\cam))e^{im\vd(v^0)} f_j(\vd(\lti)))\}.
\tag{26}
$$
But
$$
\aligned
&(\Pi \otimes I) \vd(\lti) = I \otimes \lti,\\
&(\Pi \otimes I) \vd(\cam) = M \dot{\otimes} \cam,\\
& \deka (M \dot{\otimes} \cam) = \cad_{ki}(M) \otimes \deja
(\cam),\\
&(\Pi \otimes I)(\ptak) = \ptak \otimes \ptak.
\endaligned
\tag{27}
$$
Collecting (25)--(27) we see that (24) solves the \8 condition.

The induced \6
$$
\rho_R : \cah \uparrow A(G) \to \cah \uparrow A(G) \otimes A(G)
\tag{28}
$$
is now defined in the same way as in the classical case [6]:
$$
\rho_R = I \otimes \vd.
\tag{29}
$$
For $F$ given by eq. (24) we have
$$
\rho_R(F) = (I \otimes \vd)(F) = e_i \otimes \{\deja(\vd(\cam))
e^{im(\vl^0_\mu \otimes v^\mu + v^0 \otimes I)}  f_j(\vd(\lti))\}.
\tag{30}
$$
Now, according to eqs. (22)--(24) we can identify our space of \6
with linear space of functions of $p_\mu$. The $p_\mu$'s \5e among themselves
(although they {\it do not } \5e with $v^\mu$); therefore, the
above functions can be viewed as {\it classical } functions
defined on hyperboloid $p^2 = m^2$, $p_0 > 0$, exactly as in the
classical case.

In order to simplify eq. (30) let us note first that
$$
\vd(p_\mu) = \vd(m \vl^0_\mu) = m\vl^0_\nu \otimes \vl^\nu_\mu =
p_\nu \otimes \vl^\nu_\mu,
\tag{31}
$$
so the action on the support space is standard. Next, let us
write eq. (30) in the form
$$
\aligned
\rho_R(F) & = e_i \otimes \{(\cad_{ik}(\cam) \ptak \otimes I)
(\sowa \otimes I) (\cad^{-1}_{kl}(\cam) \otimes I)\}\\
& \cdot \deja (\vd(\cam)) (\ptak \otimes I) (\sowa \otimes I)
e^{i(p_\mu \otimes v^\mu + mv^0 \otimes I)} f_j(p_\mu \otimes
\vl^\mu_\nu).
\endaligned
\tag{32}
$$
Consider the expression $(\cad^{-1}_{kl}(\cam) \otimes I)
\cad_{lj}(\vd(\cam) )$; using eqs. (21)--(23) we can at once
identify this expression. It is simply equal to $\deka
(R(p,\vl))$, {\it where } $R(p,\vl)$ {\it is a classical Wigner
rotation corresponding to the momentum } $p$ {\it and
transformation } $\vl$; of course, $\deka (R(p,\vl))$ is to be
understood here as an element of the tensor product of the algebra
of functions on the hyperboloid $p^2 = m^2$ and the algebra
$A(G)$.

As a next step we calculate
$$
(\sowa \otimes I) \deka (R(p,\vl)) (\ptak \otimes I).
$$
Using the \5ation rules among $v^0$ and $p_\mu \sim \vl^0_\mu$ we
 easily obtain that:
$$
(\sowa \otimes I) \deka (R(p,\vl)) (\ptak \otimes I). = \deka
(R(\pti,\vl)) \tag{33}
$$
where
$$\align
\pti_0 & = \frac{p_0 \cos h \mak + m \sin h \mak}{m \cos h \mak +p_0  \sin h
\mak},\\
\pti_k & = \frac{p_k}{m \cos h \mak + p_0 \sin h \mak}.
\endalign
$$
Finally,
$$
\aligned
e^{-imv^0\otimes I} & e^{i(p_\mu \otimes v^\mu + mv^0 \otimes
I)}\\
& = \exp \left\{i\kappa \ln \left(\cos h \mak + \frac{p_0}{m} \sin
h\mak\right) \otimes v^0 \right\}\\
& \cdot \exp\left\{\frac{i\kappa}{m} \frac{\sin h\mak}{\cos h
\mak + \frac{p_0}{m} \sin h \mak } p_k \otimes v^k   \right\}.
\endaligned
\tag{34}
$$
So, collecting all terms in eq. (32) we finally arrive at the
following form of our \6:
$$
\aligned
\rho_R : f_i(p) & \to\deja (R(\pti,\vl))
\exp\left\{i\kappa \ln \left( \cos h
\mak + \frac{p_0}{m} \sin h \mak\right) \otimes v^0   \right\}\\
&\exp\left\{i\kappa \frac{\sin h\mak p_k}{m\cos h
\mak + p_0 \sin h \mak } \otimes v^k \right\} f_j(p_\mu \otimes
\vl^\mu).
\endaligned
\tag{35}
$$

\head IV. Conclusions
\endhead

Let us conclude with few remarks:
\roster
\item"{(i)}" the carrier space of the \6 is, as in the classical
case, the Hilbert space of square integrable (with respect to the
standard measure $\frac{d^3\vec p}{p_0}$) functions on
hyperboloid $p^2 = m^2$;
\item"{(ii)}" in the limit $\kappa \to \infty$ one recovers the
classical \6;
\item"{(iii)}" one can pose the problem of the infinitesimal \6s,
i.e. the \6s of the `Lie algebra'; the Lie algebra of a \3 \4 is
defined via a notion of duality of Hopf algebras. For the case of
\3 $E(2)$ \4 this duality is fully understood [8], in this case,
we have checked that, using the standard definition of
infinitesimal transformations [9], one recovers the \6 of Lie
algebra from the induced \6 of $E(2)$;
\item"{(iv)}" there are many interesting questions to be studied.
First of all, as we have noted above, there are no \3 sub\4s
corresponding to light-like or space-like momenta. In the
classical case we have to use all orbits in the group dual to the
abelian sub\4 of translations in order to exhaust all irreducible
\6s. So question arises what are the \3 counterparts of mass zero
and imaginary mass \6s;
\item"{(v)}" one can define a non\5ative Minkowski space for
which the $\kappa$-\1 is the symmetry \4. One can ask what is a
\3 counterpart of classical Fourier transform relating
Poincare - covariant states with covariant space-time functions.
For example, whether there exists a \9ization of Weinberg's
theory [10].
\endroster

All the above questions will be addressed to in subsequent
publications.

{\bf Acknowledgment.}
\flushpar
I would like to thank Prof. P. Kosi/nski for available
discussions.

\enddocument